\documentclass[usenatbib]{mn2e}

\usepackage{graphicx}
\usepackage{pstricks}
\usepackage{textcomp}
\usepackage{pst-plot}
\usepackage{amssymb,amsmath} 

\newcommand{\xicl}{\xi_\mathrm{cl}}
\newcommand{\Mcl}{M_\mathrm{cl}}

\newlength{\figurewidth}
\title[A non-universal ICMF]{The galactocentric radius dependent upper mass limit of young star clusters: stochastic star formation ruled out}
\author[J.~Pflamm-Altenburg, Rosa A. Gonzalez-L\'opezlira \& Pavel Kroupa]
{Jan~Pflamm-Altenburg$^1$\thanks{email: 
    jpflamm@astro.uni-bonn.de,
    r.gonzalez@crya.unam.mx, pavel@astro.uni-bonn.de},
  Rosa A. Gonzalez-L\'opezlira$^{2}$\footnotemark[1] 
  and Pavel~Kroupa$^1$\footnotemark[1]\\
  $^1$ Helmholtz-Institut f\"ur Strahlen- und Kernphysik (HISKP), 
  University of Bonn, Nussallee 14-16, D-53115 Bonn, Germany \\
  $^2$  Centro de Radioastronom\'{\i}a y Astrof\'{\i}sica, UNAM, Campus Morelia,
   Michoac\'an, M\'exico, C.P. 58089\\
}
\begin{document}
\maketitle
\begin{abstract} 
  It is widely accepted that the 
  distribution function of the masses of 
  young star clusters is universal and can be
  purely interpreted as a probability density distribution function with 
  a constant upper mass limit. As a 
  result of this picture the masses of the most-massive objects are exclusively 
  determined   by the size of the sample. Here we show, with very high 
  confidence, that  the masses of 
  the most-massive young star clusters in M33 decrease 
  with increasing galactocentric
  radius in contradiction to the expectations from a model 
  of a randomly sampled constant
  cluster mass 
  function with a constant upper mass limit. 
  Pure stochastic star formation is thereby ruled out. 
  We use this example 
  to elucidate how naive analysis of data can lead to unphysical conclusions.
\end{abstract}
\begin{keywords}
   stars: formation -- galaxies: individual: M33 -- galaxies: star clusters: general
\end{keywords}

\section{Introduction}

Two of the most fundamental distribution functions in astronomy are
the initial stellar mass function (IMF) in star clusters and 
the initial mass function of star clusters (ICMF). Observations
have shown that the form of both distribution functions, the IMF in star 
clusters \citep{kroupa2001a,kroupa2002a,kroupa2012a} and the ICMF 
\citep[e.g.][]{degrijs2003a} seem to be universal.

There is less agreement about how the high-mass regime of both functions
is populated. On the one hand the upper mass limits are treated 
to be independent of the environment 
\citep[e.g.][]{gieles2006a,gieles2009a,parker2007a,lamb2010a} 
and therefore the most-massive 
objects are determined by a pure size-of-sample effect. 
That is, the typical mass of the most-massive object that occurs in a sample
increases with the size of the sample. 
On the other hand the formation of the most-massive objects 
could naturally require appropriate physical conditions. In the case
of the IMF there is a growing amount of evidence that massive stars 
are not forming in low-mass star clusters
\citep{weidner2006a,weidner2009a,hsu2012a}. This is incompatible with the
idea that
the high-mass regime of the IMF is populated entirely randomly.
In the case of the ICMF this is less clear. But recently, \citet{larsen2009a} 
found that if the mass distribution of
young star clusters is fitted by a Schechter-type function, then the critical 
mass, $M_\mathrm{crit}$, where the ICMF turns down at the high-mass end,
is larger for star-burst galaxies than for Milky Way type
galaxies. I.e., the most-massive star clusters in star-burst galaxies 
are more massive than those in  normal disk galaxies. \citet{larsen2009a}
suggested that this may be due to
the high-pressure environment in star-burst galaxies.  

Consequently, if the gas density determines the physical upper mass
limit for star cluster formation in whole galaxies, 
then the same effect is expected to be present within disk galaxies on smaller
scales. 
As gas densities generally increase
with decreasing galactocentric distance, the most-massive star clusters 
should form predominantly in the inner regions. One might be tempted to 
leap to the conclusion that this is the result of a size-of-sample effect 
because the star formation rate density and therefore the cluster formation rate 
density in the central regions is higher than in the outer ones.

In Sec.~\ref{sec:sos} we show that the most-massive young star clusters
in M33 seem to be in agreement at first and naive 
sight with a pure size-of-sample
 effect. In Sec.~\ref{sec:m33} we demonstrate
by using bins with equal number of star clusters that the masses of 
the most-massive star clusters decrease with increasing galactocentric radius 
therewith ruling out  a purly randomly sampled constant ICMF.

We proceed strictly logically: we assume that sampling cluster masses
from the constant ICMF is stochastic and then find extremely significant
disagreement with the observational data. The stochasticity hypothesis
is therewith falsified with extremely high confidence.

\section{The size-of-sample effect} \label{sec:sos}
The initial cluster mass function (ICMF) of young star clusters, $\xicl(\Mcl)$, 
determines the number of young star clusters, $dN$, in the mass interval
$[\Mcl,\Mcl+d\Mcl]$. Observations have shown that the ICMF can be described
by a power-law, 
\begin{equation}\label{eq:xicl}
  \xicl(\Mcl) = \frac{dN}{d\Mcl} \propto \Mcl^{-\beta}\;.
\end{equation}
In M33 \citet{sharma2011a} determine a slope of $\beta=2$ for the complete 
young star cluster population in M33. 
This is in agreement with the slope
of lower-mass young embedded star clusters in the solar neighbourhood 
\citep{lada2003a} as well as with extragalactic studies 
\citep[e.g.][]{zhang1999a,degrijs2003a}. Throughout the whole
  paper, \emph{young}
  refers to the ages of the star clusters in the sample in  \citet{sharma2011a}
  which we have used in our analysis.
  The majority of these
  star clusters have ages $\lesssim 10$~Myr. A minor fraction of the star clusters have ages up to a few tens of Myr.

The ICMF is widely interpreted as a simple universal probability density 
distribution function. As a consequence the masses of the most-massive 
star clusters are determined entirely by the size-of-sample 
effect \citep[e.g.][]{gieles2006a,gieles2009a}: If only a few star clusters 
are randomly drawn from the ICMF, then it is unlikely that a very massive 
star cluster is among this small sample.  Contrary, if a very large number of 
star clusters are drawn from the same ICMF then it is very likely that a massive
star cluster is part of this set.

Before we proceed analysing the star cluster data by \citet{sharma2011a}
it should be mentioned that, due to stochastic effects, the determination
of star cluster masses based on integrated photometry 
might be questionable for low-mass star clusters 
\citep[e.g.][]{maiz-apellaniz2009a,fouesneau2010a}.
For our analysis we use the masses of the most massive star clusters,
which are not expected to be influenced strongly by stochastic IMF effects.
The low-mass star clusters only contribute to the size-of-sample of the
considered sub-set of star clusters. As these subsets are defined 
by the position of the star clusters in M33 and not by their 
masses, stochastic IMF effects on the mass determination are unimportant.

\begin{figure}
\includegraphics[width=\columnwidth]{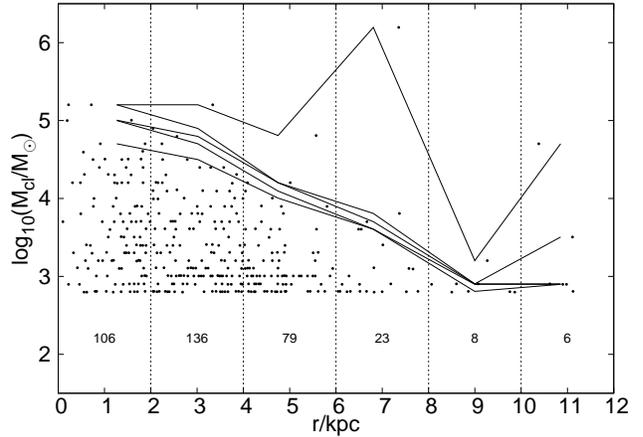}
\caption{\label{fig:m33_radial_bins} Shown are the radial position and the mass
of young star clusters (points) in M33 from \citet{sharma2011a} above 
the completion limit of $\approx 600$~M$_\odot$. The sample has been subdivided 
into radial bins with a constant width of 2~kpc. The number below the sample 
in each bin  is the number of star clusters in the bin. The solid lines 
connect the 
points indicating the 1st, 2nd, 3rd, 4th, and 5th most-massive star cluster 
in each bin. The mass of this point is the mass of the $i$-th massive object,
the radial position is the average of the radial distances of all clusters 
in the particular bin.}
\end{figure}

Figure~\ref{fig:m33_radial_bins} shows the masses versus the galactocentric
radius (black points) 
of the young star cluster sample in M33 from \citet{sharma2011a}. 
In order to allow a statistical analysis only young star clusters
above the completion limit of $\approx 600$~$M_\odot$ are considered.
The sample is divided into radial bins with a constant width of 2~kpc. 
Within each bin the 1st, 2nd, 3rd, 4th, and 5th most-massive young star cluster
is determined. The $i$-th most-massive young star cluster in a particular
bin is represented by its mass and a radial position which is the 
average of the radial positions of all star clusters in the particular bin.
The black solid lines connect the points of the $i$-th most massive star clusters
in radial direction.

It can be seen in  Fig.~\ref{fig:m33_radial_bins} that there is the general
tendency that the $i$-th most-massive star cluster becomes less massive with
increasing galactocentric radius. At the same time, 
the number of young star clusters 
in the bins (number below the star cluster sample) decreases with increasing 
galactocentric radius, too. Such a behaviour is expected if the most-massive 
star clusters are determined by the size-of-sample
effect. Theoretically, the probability density distribution of the $i$-th most-massive 
star 
cluster for a size-of-sample, $N$, is given by 
\citep[e.g.][]{pflamm-altenburg2008a}
\begin{equation}
\label{eq:ith_cmf}
\begin{array}{ccc}
  p_{i,N}(\Mcl)&=&N {N-1\choose i-1}
  \left(\int_{M_\mathrm{l}}^{M}\xicl(\Mcl)\;d\Mcl\right)^{N-i}\\
  &\times&\xicl(\Mcl)
  \left(\int_{M}^{M_\mathrm{u}}\xicl(\Mcl)\;d\Mcl\right)^{i-1}
  \end{array}\;,
\end{equation}
where $M_\mathrm{l}$ is the lower mass limit and $M_\mathrm{u}$ is the upper
mass limit of the cluster mass function and $\xicl(\Mcl)$ 
needs to be normalised, 
\begin{equation}
  \int_{M_\mathrm{l}}^{M_\mathrm{u}} \xicl(\Mcl) d\Mcl = 1\;.
\end{equation}

\begin{figure}
\includegraphics[width=\columnwidth]{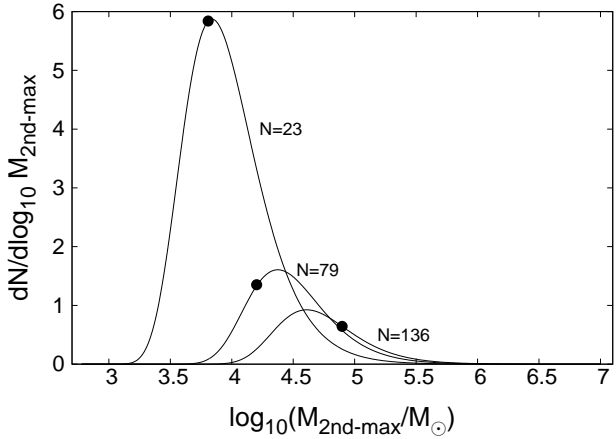}
\caption{\label{fig:2nd} The theoretical distribution of the 2nd most-massive 
  star cluster for three different sizes-of-sample: $N=23$ (6--8~kpc),
  $N=79$ (4--6~kpc), and $N=136$ (2--4~kpc). The filled circles mark the 2nd
  most-massive star cluster in the particular radial bin in M33.}
\end{figure}

Figure~\ref{fig:2nd} shows the theoretical
distribution function of the 2nd most-massive star cluster, derived from
eq.~\ref{eq:ith_cmf} with $M_\mathrm{l} = 600$~$M_\odot$ (the completness limit),
$M_\mathrm{u} = 10^{7}$~$M_\odot$, 
and $\beta = 2$ for 
three different 
sizes-of-sample ($N=136, 79, 23$, corresponding to the three radial bins
2--4~kpc, 4--6~kpc, and 6--8~kpc). The filled black circles mark 
the mass of the 2nd most-massive star cluster in the particular bin.
At first sight, the masses of the most-massive young star clusters
seem to be in agreement with the size-of-sample effect, suggesting
that there is no need for a deviation from the picture of a purely randomly
sampled universal and therefore environment independent cluster mass function.

Here, we have reached a situation that requires 
careful attention. The (naive)
conclusion that this aspect of the observations is in agreement with
the picture of  a simple universal probability  density distribution 
function of the ICMF does not imply that a different model where the formation
of the most-massive star clusters is limited by the local physical 
conditions, as for example the local gas density 
\citep{pflamm-altenburg2008a}, is ruled out. These lax dealings have lead 
to serious misconclusions. E.g., \citet{maschberger2008a} 
analyse whether the most-massive star in young low-mass star clusters is 
determined by a size-of-sample effect while drawing from a constant IMF 
with a constant upper mass limit and they state:``\emph{Our conclusion 
(in support
of the random drawing hypothisis) remains provisional}.'' But they have not
tested the alternative of a varying upper mass limit and how well such 
a  model agrees with the observations. 
Thus, careless reading of \citet{maschberger2008a} then 
leads to the acceptance of a useless statement.
E.g. \citet{eldridge2012a} refers to this work writing:
``\emph{\citet{maschberger2008a} also made 
a detailed study of all available information and also favour PSS}'' (pure-stochastic-sampling).

\section{Radial dependence}\label{sec:m33}
In order to exclude the influence of a possible size-of-sample effect
in the analysis of the star cluster masses, we  devide the sample in bins 
such that they 
contain an equal number of star clusters. If the ICMF is a simple universal 
and environment independent 
probability density distribution function, then no radial dependence of the
$i$-th most-massive star cluster is expected. 

The sample of young star clusters in M33 from \citet{sharma2011a} contains
591 clusters. For the analysis we only consider all 358 more massive than
 $600$~$M_\odot$ which is approximately the completness limit.
In order to construct as many as possible bins containing the same number of 
star clusters we choose a size-of-sample of $N=17$. This leads to only one 
redundant star cluster, which is the outer-most one. This has of course no 
effect on the statistical analysis.

\begin{figure}
  \includegraphics[width=\columnwidth]{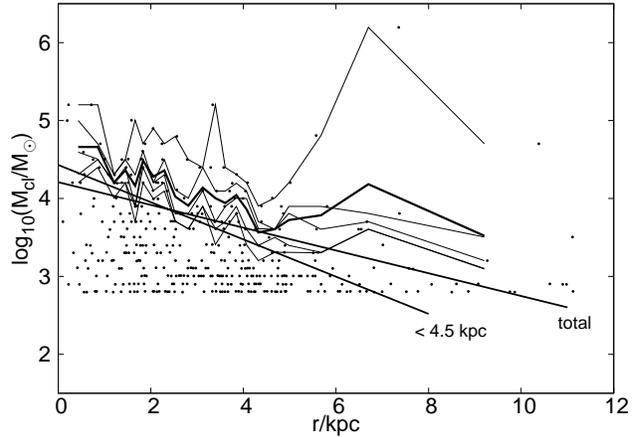}
\caption{\label{fig:M33_eqN_bins}This plot is similar to 
  Fig.~\ref{fig:m33_radial_bins}, but here the bins are chosen such that they 
  contain an equal number of star clusters ($N=17$). The 1st, 2nd, 3rd, 4th, 
  and 5th most-massive star cluster of each bin are connected by solid lines.
  The mass of this star cluster is the mass of the $i$-th massive object,
  its radial position is the average of the radial distances of all clusters 
  in the particular bin. The thick straight lines show the linear fit by 
  eq.~\ref{eq:fit} to the 5-th most-massive star cluster for two cases: (i)
  only star clusters with a galactocentric radius smaller 
  than 4.5~kpc, (ii) all 
  star clusters.}
\end{figure}

In each bin we determine the 1st to 5th most-massive star cluster. These are 
shown in Fig.~\ref{fig:M33_eqN_bins} and are connected by 
solid lines. The radial position of the $i$-th most massive star cluster 
is the average of the radial distances of all star clusters in the particular
bin. It can be seen that the trend of a decrease of the mass is preserved,
whereas no trend is expected if the upper-mass regime were purely stochastically
populated.

The radial distance dependence of the $i$-th most massive star cluster
can be described by a linear fit,
\begin{equation}\label{eq:fit}
  \log_{10}\left(\frac{\Mcl}{M_\odot}\right)
  = a\;\frac{r}{\mathrm{kpc}} + b
  \;.
\end{equation}

The coefficients $a$ and $b$ are listed in Table~\ref{tab:slopes_m33}.
Two cases are considered: i) all star clusters,
ii) only star clusters with a galactocentric distance less than 4.5~kpc.
Except for the most massive star cluster the 2nd to 5th most massive 
star cluster show a very pronounced radial dependence.

\begin{table}
  \caption{Slopes of the radial distribution of the $i$-th massive star cluster in M33}
  \label{tab:slopes_m33}
  \begin{tabular}{cccccc}
    range & $i$ &  b   &   a    & $P(\le a)$ \\ 
    \hline
    tot             &  5  & 4.209 & -0.146 &      $5.3 \times 10^{-19}$\\ 
    tot             &  4  & 4.348 & -0.158 &      $2.2 \times 10^{-16}$\\ 
    tot             &  3  & 4.465 & -0.146 &      $1.3 \times 10^{-9}$ \\ 
    tot             &  2  & 4.672 & -0.149 &      $2.5 \times 10^{-6}$\\ 
    tot             &  1  & 4.674 & -1.7$\times 10^{-4}$ &     0.52  \\ 
    \hline
    $\le$~4.5~kpc   &  5  & 4.430 & -0.239 &    $6.4 \times 10^{-11}$       \\ 
    $\le$~4.5~kpc   &  4  & 4.526 & -0.233 &    $6.6 \times 10^{-8}$        \\ 
    $\le$~4.5~kpc   &  3  & 4.583 & -0.197 &    $1.4 \times 10^{-4}$        \\ 
    $\le$~4.5~kpc   &  2  & 4.827 & -0.213 &    $1.4 \times 10^{-3}$        \\ 
    $\le$~4.5~kpc   &  1  & 5.153 & -0.208 &    $3.6 \times 10^{-2}$        \\ 
    \hline
  \end{tabular}
  
  \medskip
Listed are the coefficents $a$ and $b$ of the linear fitting function 
(eq.~\ref{eq:fit}) of the observed radial distribution of 
the 1st--5th most-massive star cluster. The last column contains 
the probability that the value of the slope $a$ or steeper in column 4 is the 
result of a randomly sampled ICMF with constant and galactocentric 
radius-independent 
upper mass limit of $M_\mathrm{u}=10^7~M_\odot$ and slope $\beta = 2$.
\end{table}

\begin{figure}
  \includegraphics[width=\columnwidth]{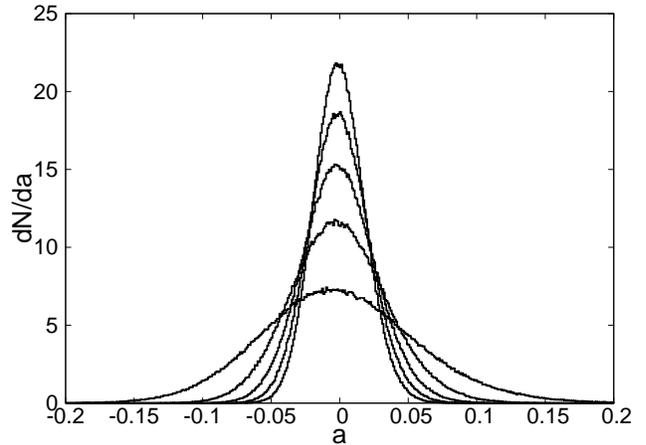}
\caption{\label{fig:M33_slopes} The distribution functions 
of the slope $a$ in the linear fitting function eq.~\ref{eq:fit} for 
the 1st, 2nd, 3rd, 4th, and 5th most-massive clusters as a result
from Monte-Carlo simulations (see Section~\ref{sec:m33} for details).}

\end{figure}

If the ICMF were universal then on average no radial trend is expected.
However, it could be possible that this snapshot of the most-massive star 
cluster sample in M33 is just the result of stochasticity of a universal
ICMF. In order to determine how likely such a radial distribution is
under the assumption of a constant and radially independent upper mass 
limit is, we 
perform a Monte-Carlo experiment: for each bin we draw $N=17$ star clusters
from an ICMF 
with $\beta=2$, $M_\mathrm{l}=600\;M_\odot$, and $M_\mathrm{up}=10^7\;
M_\odot$. The 1st to 5th most massive star clusters are identified. Their radial
positions are the radial average of the star clusters in the particular bin.
The sets are fitted by eq.~\ref{eq:fit} identically to the treatment 
of the observational data. Figure~\ref{fig:M33_slopes} shows the
distribution of the slope $a$ after 10$^6$ repetitions for case (ii), i.e. only
considering star clusters with a galactocentric distance less than 4.5~kpc.

The probability of a random drawing event of the
radial trend  is obtained from the cumulative distribution. In those cases 
where the cumulative distribution of the Monte-Carlo simulations is zero,
because 10$^6$ repetitions are too few,
the left wings of the distributions of the slope $a$ are fitted
by a Gaussian curve and extrapolated. The probabilities are
listed in Table~\ref{tab:slopes_m33}. E.g., the  observed
slope of the 3rd most-massive star cluster if all star clusters are
considered
is $a=-0.146$. The probability that  a slope of $a=-0.146$ or steeper is the 
result of a randomly sampled ICMF with a radius-independent upper mass 
limit of $10^7~M_\odot$  is
$1.3\times 10^{-9}$.

\section{parameter dependence}
\begin{figure}
  \includegraphics[width=\columnwidth]{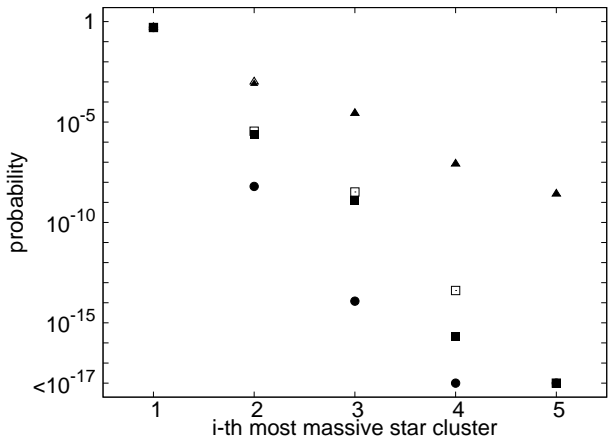}
  \caption{\label{fig:M33_probabilities} 
    The probabilities that the radial distribution of the 
    i-th most massive star cluster is the result of a randomly sampled
    constant ICMF with different upper mass limits, $M_\mathrm{u}$,  
    and ICMF-slopes, $\beta$; open symbols: $M_\mathrm{u}=10^9~M_\odot$, 
    filled symbols: $M_\mathrm{u}=10^7~M_\odot$. 
    For both upper mass limits Monte-Carlo 
    simulations for three different ICMF-slopes were performed; 
    squares: $\beta=2.0$, triangles: $\beta=1.7$, circles:
    $\beta=2.3$.}
\end{figure}

\begin{figure}
  \includegraphics[width=\columnwidth]{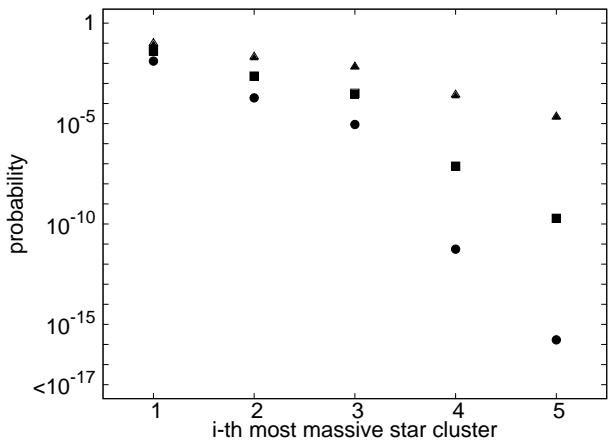}
  \caption{\label{fig:M33_probabilities_in} Same as 
    Figure~\ref{fig:M33_probabilities}, but only for the inner star clusters
    with a galactocentric distance less than 4.5~kpc.}
\end{figure}

In order to explore how the probabilities depend on the chosen parameters
of the ICMF,
we also perform Monte-Carlo simulations with different values of
the slope and the upper cluster mass limit. The probabilities are obtained
in the same way as described above. Figure~\ref{fig:M33_probabilities}
shows the probabilites for the whole set of star clusters
for an upper mass limit of $M_\mathrm{u}=10^7$~$M_\odot$
(filled symbols) and $M_\mathrm{u}=10^9$~$M_\odot$ (open symbols), and 
for three different slopes $\beta=1.7$ (triangles),  $\beta=2.0$ (squares), 
and $\beta=2.3$ (circles). Figure~\ref{fig:M33_probabilities_in} is the
same as Fig.~\ref{fig:M33_probabilities}, but only for star clusters with a 
galactocentric radius of $\le$4.5~kpc. In the cases where no open symbol appears
it lies at the same position as the corresponding filled symbol.
It can be seen that, except for the most-massive star clusters, it is unlikely
that the radial trend of the most massive star clusters is the stochastic
result of a purely randomly sampled constant ICMF. In general, the probabilities
decrease considerably with a steeper slope of the ICMF and the order of the
star cluster. 

\begin{figure}
  \includegraphics[width=\columnwidth]{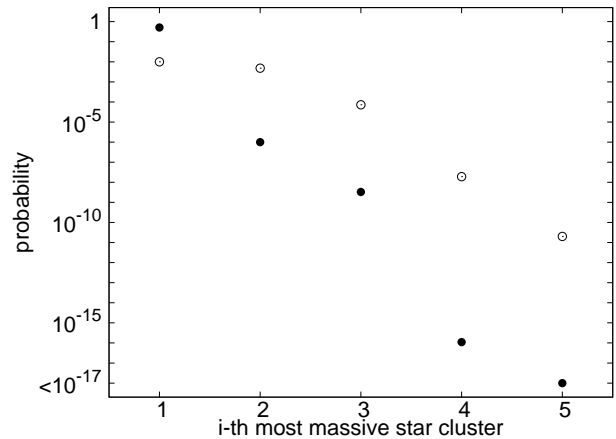}
  
  \caption{\label{fig:M33_probabilities_schechter} 
    The probabilities for a constant Schechter-type ICMF 
  with $\beta=2$ and $M_\mathrm{crit}=2.1\times 10^5 M_\odot$, as proposed  by
  \citet{larsen2009a}. The filled symbols show the probabilities for
all star clusters, the open symbols show the probabilities only for
star clusters with  a galactocentric distance of $\le$4.5~kpc.}
\end{figure}

There are several studies which conclude that the ICMF of young star clusters
may be better described by a Schechter-type function than by a single-part
power-law \citep[e.g.][]{gieles2006a,larsen2009a}. Therefore, we also test 
a Schechter-type function with a low-mass-end slope of $\beta=2$ and 
a turn-down mass of $M_\mathrm{crit}=2.1\times 10^5 M_\odot$ \citep{larsen2009a}.
The probabilities that the radial distribution of the $i$-th most massive
star cluster is consistent with a randomly sampled constant ICMF
are shown in Fig.~\ref{fig:M33_probabilities_schechter}. A constant 
randomly sampled Schechter-type ICMF is ruled out, too.

\begin{figure}
  \includegraphics[width=\columnwidth]{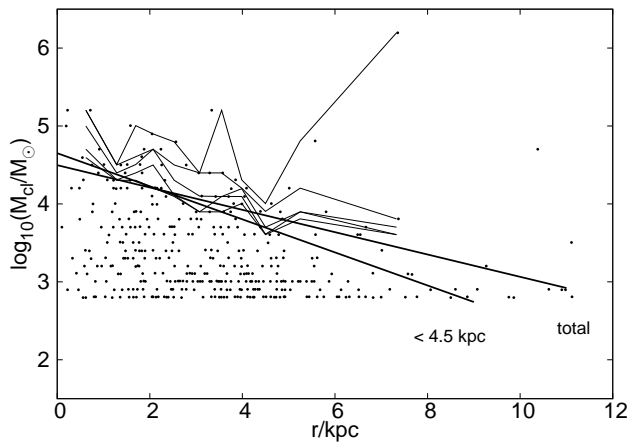}
\caption{\label{fig:M33_eqN_bins_n=32}
  This plot is similar to Fig.~\ref{fig:M33_eqN_bins} but with a size-of-sample
of $N=32$. It can bee seen that the radial trend that the masses of the most-massive star clusters decrease with increasing galacto-centric radius does not depend on the size-of-sample chosen.}
\end{figure}

\begin{figure}
 \includegraphics[width=\columnwidth]{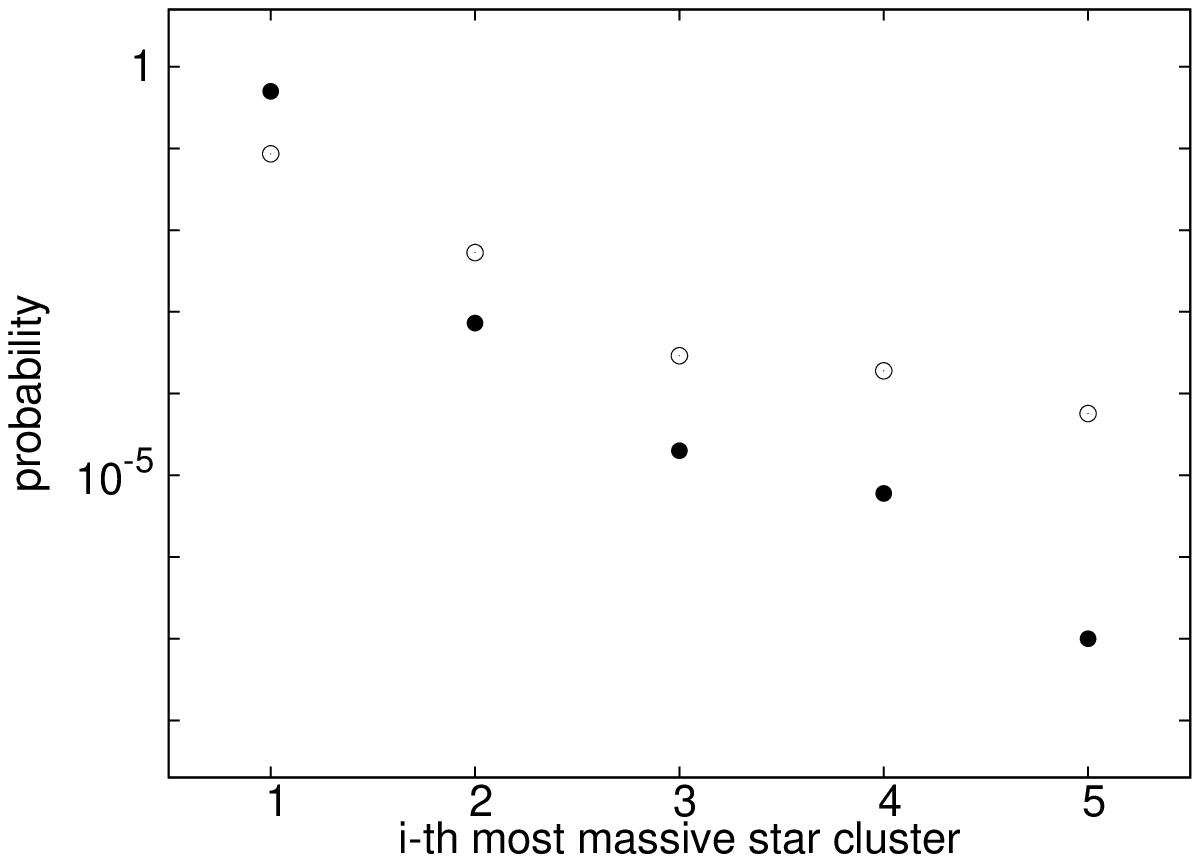}
 \caption{\label{fig:M33_probabilities_n=32} 
    The probabilities for a constant power-law ICMF 
    with $\beta=2$ and $M_\mathrm{u}=10^7 M_\odot$ but with a size-of-sample 
    of $N=32$. The filled symbols show the probabilities for
    all star clusters, the open symbols show the probabilities only for
    star clusters with  a galactocentric distance of $\le$4.5~kpc.}
\end{figure}

The analysis has been performed for a size-of-sample of $N=17$ and it might
be interesting what the results are for a different size-of-sample. 
We now set $N=32$. In this case all 358 clusters are binned 
in 11 radial bins and only 6 redundant star clusters remain. 
Fig.~\ref{fig:M33_eqN_bins_n=32} shows the radial dependence of the 
most to the 5th-most massive star cluster. It can be seen that the radial 
trend of a decrease of the masses of the most-massive star clusters 
persists. The probabilities that these trends can be the result of
a randomly sampled constant ICMF ($\beta=2$ and $M_\mathrm{u}=10^7~M_\odot$)
is shown in Fig.~\ref{fig:M33_probabilities_n=32}.

We analyse here only a radial dependence. This means that 
the observed environmental dependency of the upper mass limit of the 
ICMF must be given by the spatial location of the star clusters.
The underlying physical reason for the derived result might be that 
the surface density of the gas in M33 tendencially decreases 
with increasing galactocentric radius \citep{heyer2004a} and that
the formation of very massive star clusters require a sufficiently large
local gas reservoir. Furthermore, it is found that the masses of the
most-massive young star clusters in M33 scale with the surface gas density
\citep*{gonzalez-lopezlira2012a}. 
In this context, the existence of the very massive star cluster in the outer
region should not be an exception; the local gas density must be
high enough in  order to form it.

\begin{figure}
  \includegraphics[width=\columnwidth]{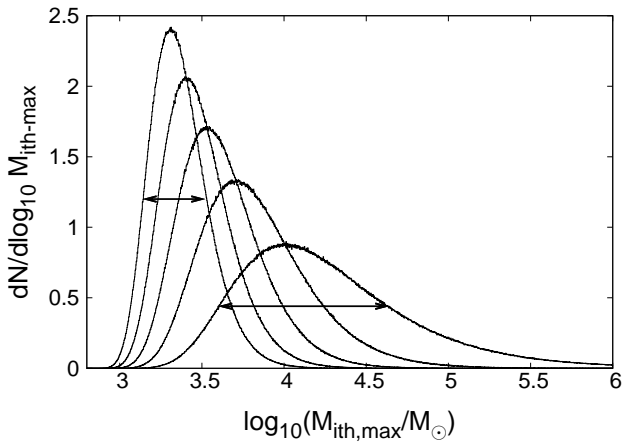}
  \caption{\label{fig:m_1_to_5_dist} The distribution of the 
    1st to 5th most-massive star cluster for a size-of-sample of $N=17$ 
    star clusters and an ICMF with slope $\beta=2$ between 
    $M_\mathrm{l}=600 M_\odot$ and 
    $M_\mathrm{u}=10^{7} M_\odot$. The arrows mark the FWHM of the distribution 
of the most-massive star cluster (1.0~dex) and of the fith 
    most-massive star cluster (0.35~dex).}
\end{figure}

The high-probability in  the case of the
most-massive star cluster does not mean that a non-universal ICMF is ruled out
but only that the consistency with a universal ICMF is much higher than for
the star clusters with a lower order. This is not surprising. 
Figure~\ref{fig:m_1_to_5_dist} shows the distribution of the 1st to 5th 
most-massive star cluster for a size-of-sample of $N=17$ star clusters and 
an ICMF with slope
$\beta=2$ between $M_\mathrm{l}=600 M_\odot$ and $M_\mathrm{u}=10^{7} M_\odot$.
The distribution of the most-massive star cluster is much broader than for
the 5th-most massive star cluster. Specifically, the FWHM for the most-massive
star cluster is 1.0~dex, whereas the FWHM for the 5th most-massive star cluster 
is only 0.35~dex. 
If there is a radial dependence of the upper mass limit of the ICMF (being
higher in the central region, i.e. in regions of higher gas densities)
then the broad 
distribution of the most-massive young star cluster 
smears out the radial trend.
This effect has also been mentioned in \citet{larsen2009a} stating that the
error is smaller for the 5th brightest star cluster than for the brightest. 
Therefore, it is more reliable to use the fith-most massive star cluster 
and thus a randomly sampled constant and environmentaly  independent ICMF 
is ruled out.

A final remark should be made if disruption processes of 
young star clusters can cause the observed trend. Disruption due to mass loss by stellar evolution and gas expulsion are
internal processes and should occur in the same way at each position in the galaxy and can therefore not create the observed trend. Disruption due to tidal effects should depend on the position of a star cluster in a galaxy. However, 
tidal effects mainly have influence on the evolution of
lower-mass star clusters and we assume that high-mass star cluster
are not disrupted on the short time scale of $\lesssim$10--20~Myr.

\section{Conclusion}

We have found that the formation of very massive star clusters 
is increasingly suppressed with increasing galactocentric radius in M33. 
We have ruled out with extremely high significance that this is the result
of a size-of-sample effect, where a constant and environment independent 
ICMF is populated entirely randomly and environmental effects can be neglected.

The straightforward conclusion is that 
very massive star clusters require special physical conditions in order
to form, as 
for example high gas surface densities. This is indeed a very
plausible notion, because gas surface densities decrease tendentially with 
increasing galactocentric radius.

If the ICMF is not independent of the environment, why should the stellar IMF
be? Indeed, direct analysis of Galactic star forming regions shows that the 
formation of high-mass stars takes place in high-mass star clusters which
can not be a size-of-sample effect
\citep{weidner2006a,weidner2009a,hsu2012a}.
The combination of both an environmently dependent ICMF and IMF is thus required
to calculate the IMF of whole galaxies by adding all individual IMFs
of all star clusters. This has been formulated in the theory of the
IGIMF (integrated galactic stellar initial mass function). The main
property of the IGIMF is that galaxies with a low star-formation rate
have steeper IGIMFs than galaxies with a high star-formation rate
\citep{weidner2003a,weidner2004b,weidner2005a}.
The non-identity of the IMF in star-clusters and in whole galaxies 
requires the revision of the calibration of star-formation tracers and 
of the calculation of star-formation rates 
\citep{pflamm-altenburg2007d,pflamm-altenburg2009a,pflamm-altenburg2009c}
and of the chemical evolution of galaxies \citep{koeppen2007a,recchi2009a}.

In this context the work presented here fundamentally supports 
the IGIMF theory and its consequences.

\bibliographystyle{mn2e}
\bibliography{m33,star-formation,cmf,imf,star-cluster,dwarf_galaxies,galaxy-evolution,OB-star}

\begin{thebibliography}{}

\bibitem[\protect\citeauthoryear{{de Grijs}, {Anders}, {Bastian}, {Lynds},
  {Lamers} \& {O'Neil}}{{de Grijs} et~al.}{2003}]{degrijs2003a}
{de Grijs} R.,  {Anders} P.,  {Bastian} N.,  {Lynds} R.,  {Lamers}
  H.~J.~G.~L.~M.,    {O'Neil} E.~J.,  2003, \mnras, 343, 1285

\bibitem[\protect\citeauthoryear{{Eldridge}}{{Eldridge}}{2012}]{eldridge2012a}
{Eldridge} J.~J.,  2012, \mnras, 422, 794

\bibitem[\protect\citeauthoryear{{Fouesneau} \& {Lan{\c c}on}}{{Fouesneau} \&
  {Lan{\c c}on}}{2010}]{fouesneau2010a}
{Fouesneau} M.,  {Lan{\c c}on} A.,  2010, \aap, 521, A22

\bibitem[\protect\citeauthoryear{{Gieles}}{{Gieles}}{2009}]{gieles2009a}
{Gieles} M.,  2009, \apss, 324, 299

\bibitem[\protect\citeauthoryear{{Gieles}, {Larsen}, {Bastian} \&
  {Stein}}{{Gieles} et~al.}{2006}]{gieles2006a}
{Gieles} M.,  {Larsen} S.~S.,  {Bastian} N.,    {Stein} I.~T.,  2006, \aap,
  450, 129

\bibitem[\protect\citeauthoryear{{Gonz{\'a}lez-L{\'o}pezlira},
  {Pflamm-Altenburg} \& {Kroupa}}{{Gonz{\'a}lez-L{\'o}pezlira}
  et~al.}{2012}]{gonzalez-lopezlira2012a}
{Gonz{\'a}lez-L{\'o}pezlira} R.~A.,  {Pflamm-Altenburg} J.,    {Kroupa} P.,
  2012, \apj, 761, 124

\bibitem[\protect\citeauthoryear{{Heyer}, {Corbelli}, {Schneider} \&
  {Young}}{{Heyer} et~al.}{2004}]{heyer2004a}
{Heyer} M.~H.,  {Corbelli} E.,  {Schneider} S.~E.,    {Young} J.~S.,  2004,
  \apj, 602, 723

\bibitem[\protect\citeauthoryear{{Hsu}, {Hartmann}, {Allen}, {Hernandez},
  {Megeath}, {Mosby}, {Tobin} \& {Espaillat}}{{Hsu} et~al.}{2012}]{hsu2012a}
{Hsu} W.-H.,  {Hartmann} L.,  {Allen} L.,  {Hernandez} J.,  {Megeath} S.~T.,
  {Mosby} G.,  {Tobin} J.~J.,    {Espaillat} C.,  2012, ArXiv e-prints

\bibitem[\protect\citeauthoryear{{K{\"o}ppen}, {Weidner} \&
  {Kroupa}}{{K{\"o}ppen} et~al.}{2007}]{koeppen2007a}
{K{\"o}ppen} J.,  {Weidner} C.,    {Kroupa} P.,  2007, \mnras, 375, 673

\bibitem[\protect\citeauthoryear{{Kroupa}}{{Kroupa}}{2001}]{kroupa2001a}
{Kroupa} P.,  2001, \mnras, 322, 231

\bibitem[\protect\citeauthoryear{{Kroupa}}{{Kroupa}}{2002}]{kroupa2002a}
{Kroupa} P.,  2002, \sci, 295, 82

\bibitem[\protect\citeauthoryear{{Kroupa} \& {Weidner}}{{Kroupa} \&
  {Weidner}}{2003}]{weidner2003a}
{Kroupa} P.,  {Weidner} C.,  2003, \apj, 598, 1076

\bibitem[\protect\citeauthoryear{{Kroupa}, {Weidner}, {Pflamm-Altenburg},
  {Thies}, {Dabringhausen}, {Marks} \& {Maschberger}}{{Kroupa}
  et~al.}{2011}]{kroupa2012a}
{Kroupa} P.,  {Weidner} C.,  {Pflamm-Altenburg} J.,  {Thies} I.,
  {Dabringhausen} J.,  {Marks} M.,    {Maschberger} T.,  2011, ArXiv e-prints

\bibitem[\protect\citeauthoryear{{Lada} \& {Lada}}{{Lada} \&
  {Lada}}{2003}]{lada2003a}
{Lada} C.~J.,  {Lada} E.~A.,  2003, \araa, 41, 57

\bibitem[\protect\citeauthoryear{{Lamb}, {Oey}, {Werk} \& {Ingleby}}{{Lamb}
  et~al.}{2010}]{lamb2010a}
{Lamb} J.~B.,  {Oey} M.~S.,  {Werk} J.~K.,    {Ingleby} L.~D.,  2010, \apj,
  725, 1886

\bibitem[\protect\citeauthoryear{{Larsen}}{{Larsen}}{2009}]{larsen2009a}
{Larsen} S.~S.,  2009, \aap, 494, 539

\bibitem[\protect\citeauthoryear{{Ma{\'{\i}}z Apell{\'a}niz}}{{Ma{\'{\i}}z
  Apell{\'a}niz}}{2009}]{maiz-apellaniz2009a}
{Ma{\'{\i}}z Apell{\'a}niz} J.,  2009, \apj, 699, 1938

\bibitem[\protect\citeauthoryear{{Maschberger} \& {Clarke}}{{Maschberger} \&
  {Clarke}}{2008}]{maschberger2008a}
{Maschberger} T.,  {Clarke} C.~J.,  2008, \mnras, 391, 711

\bibitem[\protect\citeauthoryear{{Parker} \& {Goodwin}}{{Parker} \&
  {Goodwin}}{2007}]{parker2007a}
{Parker} R.~J.,  {Goodwin} S.~P.,  2007, \mnras, 380, 1271

\bibitem[\protect\citeauthoryear{{Pflamm-Altenburg} \&
  {Kroupa}}{{Pflamm-Altenburg} \& {Kroupa}}{2008}]{pflamm-altenburg2008a}
{Pflamm-Altenburg} J.,  {Kroupa} P.,  2008, \nat, 455, 641

\bibitem[\protect\citeauthoryear{{Pflamm-Altenburg} \&
  {Kroupa}}{{Pflamm-Altenburg} \& {Kroupa}}{2009}]{pflamm-altenburg2009c}
{Pflamm-Altenburg} J.,  {Kroupa} P.,  2009, \apj, 706, 516

\bibitem[\protect\citeauthoryear{{Pflamm-Altenburg}, {Weidner} \&
  {Kroupa}}{{Pflamm-Altenburg} et~al.}{2007}]{pflamm-altenburg2007d}
{Pflamm-Altenburg} J.,  {Weidner} C.,    {Kroupa} P.,  2007, \apj, 671, 1550

\bibitem[\protect\citeauthoryear{{Pflamm-Altenburg}, {Weidner} \&
  {Kroupa}}{{Pflamm-Altenburg} et~al.}{2009}]{pflamm-altenburg2009a}
{Pflamm-Altenburg} J.,  {Weidner} C.,    {Kroupa} P.,  2009, \mnras, 395, 394

\bibitem[\protect\citeauthoryear{{Recchi}, {Calura} \& {Kroupa}}{{Recchi}
  et~al.}{2009}]{recchi2009a}
{Recchi} S.,  {Calura} F.,    {Kroupa} P.,  2009, \aap, 499, 711

\bibitem[\protect\citeauthoryear{{Sharma}, {Corbelli}, {Giovanardi}, {Hunt} \&
  {Palla}}{{Sharma} et~al.}{2011}]{sharma2011a}
{Sharma} S.,  {Corbelli} E.,  {Giovanardi} C.,  {Hunt} L.~K.,    {Palla} F.,
  2011, \aap, 534, A96

\bibitem[\protect\citeauthoryear{{Weidner} \& {Kroupa}}{{Weidner} \&
  {Kroupa}}{2005}]{weidner2005a}
{Weidner} C.,  {Kroupa} P.,  2005, \apj, 625, 754

\bibitem[\protect\citeauthoryear{{Weidner} \& {Kroupa}}{{Weidner} \&
  {Kroupa}}{2006}]{weidner2006a}
{Weidner} C.,  {Kroupa} P.,  2006, \mnras, 365, 1333

\bibitem[\protect\citeauthoryear{{Weidner}, {Kroupa} \& {Bonnell}}{{Weidner}
  et~al.}{2010}]{weidner2009a}
{Weidner} C.,  {Kroupa} P.,    {Bonnell} I.~A.~D.,  2010, \mnras, 401, 275

\bibitem[\protect\citeauthoryear{{Weidner}, {Kroupa} \& {Larsen}}{{Weidner}
  et~al.}{2004}]{weidner2004b}
{Weidner} C.,  {Kroupa} P.,    {Larsen} S.~S.,  2004, \mnras, 350, 1503

\bibitem[\protect\citeauthoryear{{Zhang} \& {Fall}}{{Zhang} \&
  {Fall}}{1999}]{zhang1999a}
{Zhang} Q.,  {Fall} S.~M.,  1999, \apjl, 527, L81

\end{thebibliography}

\end{document}